\documentclass[aps,twocolumn,showpacs,preprintnumbers,amsmath,groupedaddress,superscriptaddress,amssymb,pre]{revtex4-1}
\usepackage{epsf,amsmath,amssymb,verbatim,color,multirow,pifont}
\usepackage{graphicx}

\begin{document}
\title{Crossover behavior of conductivity in a discontinuous percolation model}

\author{Seongmin Kim}
\affiliation{Department of Physics and Astronomy, Seoul National University 151-747, Korea}
\author{Y.S. Cho}
\affiliation{Department of Physics and Astronomy, Seoul National University 151-747, Korea}
\author{N.A.M.  Ara\'ujo}
\affiliation{Computational Physics for Engineering Materials, IfB, ETH Zurich, Wolfgang-Pauli-Strasse 7, CH-8093 Zurich, Switzerland}
\author{B. Kahng}
\affiliation{Department of Physics and Astronomy, Seoul National University 151-747, Korea}
\email{bkahng@snu.ac.kr}

\begin{abstract}
When conducting bonds are occupied randomly in a two-dimensional square
lattice, the conductivity of the system increases continuously as the
density of those conducting bonds exceeds the  percolation threshold.
Such a behavior is well known in percolation theory; however, the
conductivity behavior has not been studied yet when the percolation
transition is discontinuous. Here we investigate the conductivity
behavior through a discontinuous percolation model evolving under a
suppressive external bias. Using effective medium theory, we
analytically calculate the conductivity behavior as a function of the
density of conducting bonds. The conductivity function exhibits a
crossover behavior from a drastically to a smoothly increasing function
beyond the percolation threshold in the thermodynamic limit. The
analytic expression fits well our simulation data.   
\end{abstract}

\pacs{64.60.ah,02.50.Ey,89.75.Hc}
 
\maketitle

The concept of percolation transition has played a central role as a
model for the formation of a spanning cluster connecting two opposite
edges of a system in Euclidean space as a control parameter $p$ is
increased beyond a certain threshold $p_c$ \cite{stauffer}.  This model
has been used to study many phenomena such as metal-insulator
transitions and sol-gel transitions.  The order parameter $P_{\infty}$
of percolation transition is defined as the probability that a bond
belongs to a spanning cluster, which increases in the form
$P_{\infty}(p)\sim (p-p_c)^{\beta}$ beyond $p_c$, where $p$ is a control
parameter indicating the fraction of occupied bonds and $\beta$ is the
critical exponent related to the order parameter. As an application of
percolation model, one can construct a random resistor network in which
each occupied bond is regarded as a resistor with unit resistance, and
the system is in contact with two bus bars at the opposite edges of the
system.  When a voltage difference is applied between these two bus bars,
the system is in a insulating state for $p<p_c$, but changes to
conducting state for $p>p_c$, due to the formation of several conducting
paths at $p_c$. Above $p_c$, the conductivity increases continuously as
$g \sim (p-p_c)^{\mu}$, where $\mu$ is the conductivity exponent
\cite{pose}.

\begin{figure*}[t]
\includegraphics[width=1.0\linewidth]{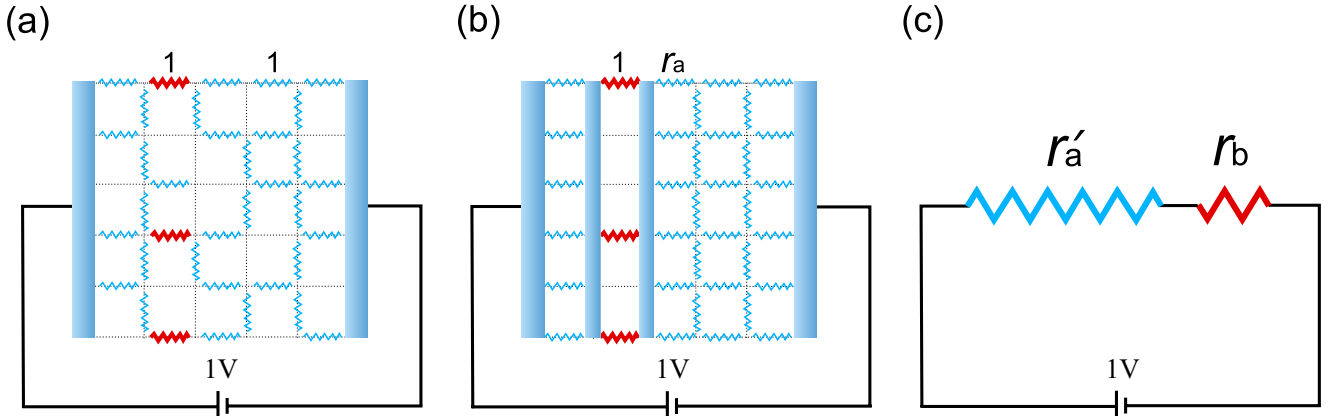}
\caption{(Color online) (a) Schematic diagram of circuit structure for
$p \geq p_{cm}$ for the SCA model that consists of bonds of unit resistance.
The occupied bonds are classified into original bridge bonds (thick red resistors) and original non-bridge bonds 
(thin blue resistors). (b) We simplify the whole circuit as series connection of a bundle of
original bridge bonds of unit resistance and two compact clusters consist of bonds of resistance $r_a$ by applying effective medium theory. (c) The combined resistance of two compact clusters is calculated as $r_a^{\prime}=r_a(L-1)/L\approx r_a$ for large $L$, and the combined resistance of original bridge bonds $r_b$ is calculated as 
$r_b = 1/Lp_b$, where the derivations are shown in the main text. We can calculate the conductivity 
as $g_m(p)=1/(r_a^{\prime}+r_b)$. \label{fig::fig1}
 }
\end{figure*}

\begin{figure}[b]
\includegraphics[width=1.0\linewidth]{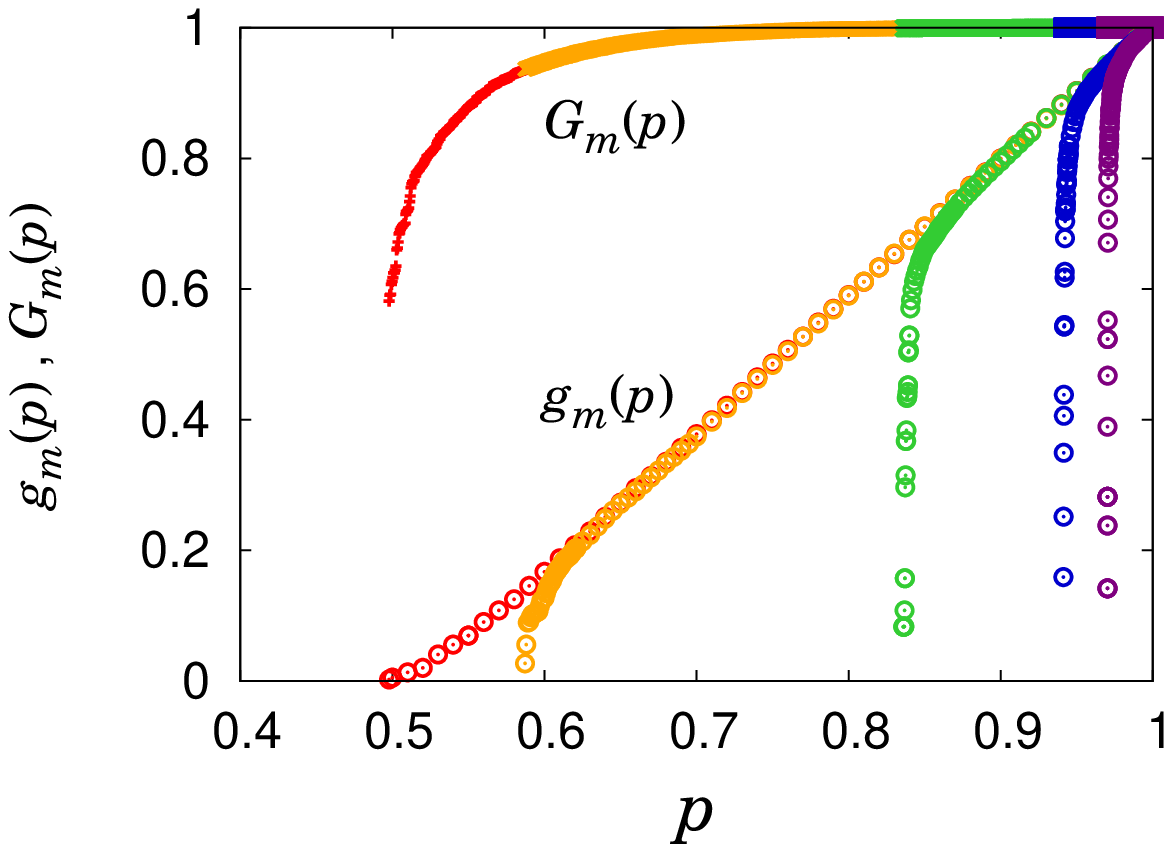}
\caption{(Color online) Plot of $G_m(p)$ and $g_m(p)$ vs $p$. $G_m(p)$ is the
fraction of nodes belonging to the spanning cluster. 
$G_m(p)$ jumps to $G_1(p)$ from 0 at $p_{cm}$ and follows the envelop of $G_1(p)$ after that.
$g_m(p)$ is the conductivity, it becomes positive at $p_{cm}$ and grows drastically after that.
As $m$ is increased, $p_{cm}$ is delayed. Data  are shown for $m=1$ (red), $m=2$ (yellow), 
$m=3$ (green), $m=4$ (blue), and $m=5$ (purple) from left to right. $L=300$ are considered. 
Results are for a single sample.
\label{fig::fig2}}
\end{figure}

Recently the subject of discontinuous percolation transition (DPT) has
been a central
issue~\cite{explosive,ziff,moreira,cho1,cho2,cho3,araujotric,andrade,costa,riordan}
with, for example, applicability to cascading failures in complex
networks~\cite{havlin}. Among
others~\cite{araujo,schrenk,nagler,bfw1,bfw2,intra}, a model called spanning
cluster avoiding (SCA) was introduced~\cite{sca} aiming to generate a
DPT. The DPT of the SCA  model is rather trivial, for the percolation
threshold is placed at $p_c=1$ in the thermodynamic limit, but for
finite-sized systems $p_c<1$. Here, we study the conductivity as a
function of $p$ in finite-sized systems for the SCA model. Indeed, we
find that the conductivity increases drastically just after the
percolation threshold and then exhibits a crossover  to a smoothly
increasing behavior.   Such crossover has never been reported, though it
is meaningful, as, a drastic change of conductivity in random resistor
networks can find application, for example, on resistance switching
phenomena in non-volatile memory devices~\cite{nonvolite}. From a
theoretical perspective, the understanding of conductivity becomes
complementary to the result on the percolation transition for the SCA
model. 

We first recall the SCA model. In this model, we take a two-dimensional
regular square lattice of linear size $L$. Initially, the system
consists of $N=L^2$ nodes and $2N$ unoccupied bonds. At each time step,
one randomly chooses $m$ unoccupied bonds, and  those potential bonds
are classified into two types: bridge and non-bridge bonds. Bridge bonds
are those that would form a spanning cluster if any of them is
occupied~\cite{hans,hans2}. One takes a non-bridge bond randomly among those
$m$ candidates if exists.  This choice suppresses the formation of a
spanning cluster.  As the number of occupied bonds is increased, the
total number of bridge bonds $N_{BB}(p)$ increases and thus the
probability that those $m$ bonds are all bridge bonds is also increased.
If such a case happens, a bridge bond is inevitably occupied and a
spanning cluster is formed. Once a spanning cluster is formed, no more
restrictions are imposed on the occupation of bonds. It was found that
when $m$ is greater than a tricritical point $m_c\approx 2.55$ in two dimensions, the percolation
transition is discontinuous and the percolation threshold $p_{cm}$
approaches unity as the system size is increased~\cite{sca}.  In
finite-sized systems, the percolation threshold $p_{cm}$ depends on the
number of candidate bonds $m$. In this brief report, we present an
analytic formula for the conductivity based on effective medium
theory~\cite{rmp}.  The analytic prediction of the conductivity function
is in agreement with our numerical data. 

In the SCA model, the percolation threshold is delayed by suppressing
the formation of a spanning cluster. While the percolation threshold is
delayed, two large clusters form independently, which are separated by
bridge bonds. Bridge bonds form a fractal set of fractal dimension
$d_{\rm BB}\approx 1.215$~\cite{hans,hans}.  Moreover, for $m>m_c$, in those
two separated clusters, the density of occupied bonds is extremely high,
for $p$ close to and above $p_{cm}$. These facts enable us to apply
effective medium theory to calculate the conductivity function near the
percolation threshold for the SCA  model. We recall the conductivity
function for ordinary percolation obtained from effective medium theory
near $p=1$, which is $g_{\rm eff}=2p-1$ in two dimensions \cite{rmp}.  

Next, we derive a formula for the conductivity using heuristic
arguments. To proceed, we examine the structure of the system at the
onset of the percolation transition, denoted as $p_{cm}^-$. As shown in
Fig.~\ref{fig::fig1}(a), the system consists of  two disconnected
clusters separated by unoccupied bridge bonds. Due to the unoccupied
separatrix, the conductivity of the system is zero at $p_{cm}^-$, but
becomes nonzero once a bond among those bridge bonds is occupied as shown
in the Fig.~\ref{fig::fig2}.  Since the number of bridge bonds increases
as $p$ increases for $p < p_{cm}$, from now on, we use the phrase
``original bridge bonds'' to refer to those bonds that were bridge bonds
at $p_{cm}^{-}$. Similarly, ``original non-bridge bonds'' are all the
other bonds.  The densities of occupied bonds of original non-bridge
bonds and original bridge bonds are denoted as $p_a$ and $p_b$,
respectively. Those two densities depend on $p$, and $p_b=0$ at
$p_{cm}^-$. Then, the following relation holds,
\begin{equation}
2L^2p \approx p_a (2L^2-L^{d_{\rm BB}})+p_bL^{d_{\rm BB}}\approx 2L^2p_a
+p_bL^{d_{\rm BB}},
\end{equation}     
where we use the number of bridge bonds $N_{BB}(p_{cm}^{-}) \approx L^{d_{BB}}$ and $L^2 \gg L^{d_{\rm
BB}}$ for large $L$. Then $p_a\approx p+{\cal O}(1/L^{2-d_{\rm BB}})$.
For $p_b(p)$, we use the fact that the occupation of original bridge
bonds increases linearly with increasing $p$ for $p > p_{cm}$ and
$p_b(p_{cm})=0$. Then, one obtains,
\begin{equation} 
p_b(p)=\frac{p-p_{cm}}{1-p_{cm}}.
\end{equation}

In the spirit of effective medium theory, we assume that original
non-bridge bonds are fully occupied but we consider that each bond has
resistance $r_a\ne 1$.  Next, we make a more crude assumption. Due to
the fractal nature of the set of bridge bonds, the separatrix is not
linear in its shape, and the density of original non-bridge and original
bridge bonds are different. Thereby, the current can flow along the
boundary between the original non-bridge and original bridge bonds.
However, this current contribution to the conductivity of the system can
be negligible when the system size is sufficiently large. Based on such
facts, we simplify the system as shown in
Figs.~\ref{fig::fig1}(b)~and~(c). That is, the system consists of two
parts, a rectangular-shape regular lattice of size $(L-1)\times L$ in
which original all non-bridge bonds are all occupied with resistance $r_a=2p_a-1$, and 
one dimensional columnar lattice of size $L$ in
which original bridge bonds are occupied with probability $p_b$  and
unit resistance. We also assume that there exists a busbar between the
two parts, and thereby there is no net current on each vertical bond.  
 
\begin{figure}[b]
\includegraphics[width=1.0\linewidth]{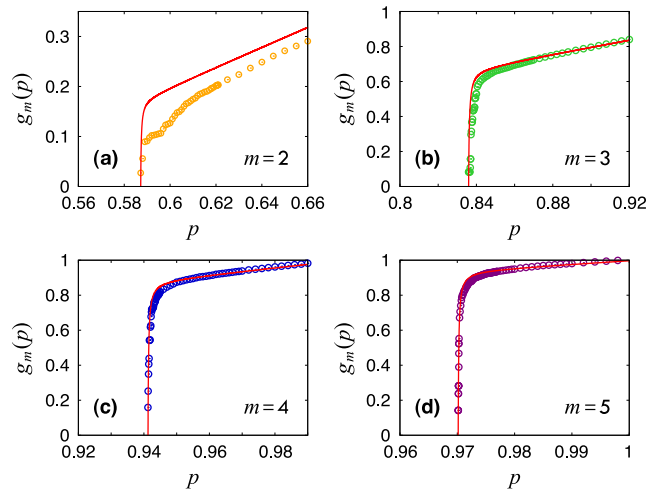}
\caption{(Color online) Plot of $g_m(p)$ vs $p$ for one sample with $L=300$. Just after $p_{cm}$,
$g_m(p)$ becomes positive and grows drastically.
Red solid lines are obtained from Eq.~(3) for $m=2$ (a), $m=3$ (b), $m=4$ (c), and $m=5$ (d).
We can find that the theoretical formula fits well the simulation data when 
$m$ is larger than the tricritical point $m_c \approx 2.55$. \label{fig::fig3}
}
\end{figure}

This simplified picture enables us to calculate the
overall conductivity. The resistivity (the inverse of
conductivity) is obtained as 
\begin{equation}
\frac{1}{g_{m}(p)}\approx \frac{1}{2p_a-1}+\frac{1}{Lp_b}, 
\end{equation}
where $g_m(p)$ denotes the conductivity at $p$ of the SCA model with the
control parameter $m$.  We compare the analytic result with our
simulation data for different $m=2,3,4$, and 5. As can be seen in
Fig.~\ref{fig::fig3}, the data for $L\times L=300\times 300$ is in good
agreement with the analytic expression for $m \ge 3$.  Since for these
cases $p_{cm}$ is close to unity (for example, $p_{c3}\approx 0.84$,
$p_{c4}\approx 0.94$, and $p_{c5}\approx 0.97$), the approximation based
on effective medium theory is more accurate.  For $m=2$, the data
clearly differs from the analytic expression. Actually, the percolation
threshold for $m=2$ reduces to the one of the ordinary
percolation in the thermodynamic limit and, therefore, the two
clusters connected through original bridge bonds cannot be considered
compact, as necessary to apply effective medium theory.

Finally, we recall the previous result~\cite{sca} that $p_{cm}$
approaches to one as $L$ is increased as 
\begin{equation}
1-p_{cm}\sim L^{-\frac{2}{m-1}(\frac{m}{m_c}-1)}~~{\rm for}~~m>m_c,
\end{equation}
where $m_c \approx 2.55$ in two dimensions. Then, $Lp_b\equiv
L^{\alpha}(p-p_{cm})$, where 
\begin{equation}
\alpha=1+\frac{2}{m-1}\Big(\frac{m}{m_c}-1\Big).
\end{equation}
Numerically, $\alpha\approx 1.18$, 1.38, and 1.48 for $m=3,4$, and $5$ in
two dimensions, respectively.  Depending on the magnitude of $p-p_{cm}$,
the conductivity behaves as follows,
\begin{displaymath}
g_m(p) \approx \left\{ \begin{array}{ll} L^{\alpha}(p-p_{cm}) &
\textrm{for $\delta \ll 1/L^{\alpha}$}\\
2p-1 & \textrm{for $\delta \gg 1/L^{\alpha}$},
\end{array} \right.
\end{displaymath}
where $\delta=p-p_{cm}$.  Thus, there exists a crossover in the
conductivity for $\delta_c \approx 1/L^{\alpha}$. We remark that the
conductivity increases more rapidly to $2p_{cm}-1$ for larger systems
due to the prefactor $L^{\alpha}$. 

In summary, we studied the conductivity transition of two dimensional
SCA model.  In this model, $p_{cm}$ increases to $1$ for $m>m_c \approx
2.55$, but otherwise it decreases to $p_{c1}=0.5$ as the system size
increases. We used effective medium theory which is valid for $p \gg
p_{c1}$ to calculate the analytic expression of conductivity in this model.
We numerically confirmed the validity of this expression for $m=2,3,4$, and
$5$ in finite sized systems and found that the data is well fitted for
$m=3,4$, and $5$. However, the case $m=2$ cannot be described by our
theory.

This work was supported by the NRF grants (Grant No.2010-0015066) and the Global Frontier Program (YSC).

\vfil\eject
\end{document}